\documentclass[preprint,3p]{elsarticle}
\usepackage{amssymb}
\usepackage{amsthm}
\usepackage{graphicx}
\usepackage{units}
\usepackage{url}
\usepackage{amsmath}
\usepackage{amsfonts}
\usepackage{bm}
\usepackage{textcomp}
\usepackage{subfigure}
\usepackage{multicol}
\usepackage{multirow}
\usepackage{verbatim}
\usepackage{rotating}
\usepackage[colorlinks,linkcolor=black,citecolor=red]{hyperref}
\usepackage{float}
\usepackage{epsfig}
\usepackage{dcolumn}
\usepackage{bm}
\usepackage{color}
\usepackage{pstricks}
\usepackage{pst-node}
\usepackage{times}
\usepackage{indentfirst}
\usepackage[english]{babel}
\usepackage{epstopdf}

\ifluatex 
    \usepackage{fontspec}
    \defaultfontfeatures{Ligatures={TeX}}
\else
    \usepackage[utf8]{inputenc}
    \usepackage[T2A,T1]{fontenc}
\fi

\addto{\captionsenglish}{%

}
\journal{Physics Letters B}
\makeatletter

\newenvironment{figurehere}
  {\def\@captype{figure}}
  {}
\makeatother

\usepackage[font=footnotesize]{caption}

\newcommand{\jpsi}{$J/\psi$~}

\newcommand{\eetoee}{$e^+e^- \to e^+e^-$~}
\newcommand{\eetomumu}{$e^+e^- \to \mu^+\mu^-$~}
\newcommand{\eetohad}{$e^+e^- \to \rm inclusive \  \rm hadrons$~}
\newcommand{\eetodigam}{$e^+e^- \to \gamma\gamma$~}
\newcommand{\eetotwogam}{$\gamma^{*}\gamma^{*} \to X$~}

\newcommand{\eetomumugam}{$e^+e^- \to \gamma^{\rm ISR}\mu^+\mu^-$~}

\newcommand{\eetomumuwos}{$e^+e^- \to \mu^+\mu^-$}

\newcommand{\Bjpsitoll}{B(J/\psi \to l^+l^-)}

\newcommand{\Tw}{\Gamma_{\rm tot}}
\newcommand{\Ew}{\Gamma_{ee}}
\newcommand{\Mw}{\Gamma_{\mu\mu}}
\newcommand{\Lw}{\Gamma_{ll}}

\newcommand{\EwDMw}{\Gamma_{ee}/\Gamma_{\mu\mu}}
\newcommand{\LwDTw}{\Gamma_{ll}/\Gamma_{\rm tot}}
\newcommand{\EwMEwDTw}{\Gamma_{ee}\Gamma_{ee}/\Gamma_{\rm tot}}
\newcommand{\EwMMwDTw}{\Gamma_{ee}\Gamma_{\mu\mu}/\Gamma_{\rm tot}}
\newcommand{\LwMLwDTw}{\Gamma_{ll}\Gamma_{ll}/\Gamma_{\rm tot}}

\newcommand{\csth}{\cos\theta}

\biboptions{numbers,sort&compress}

\begin{document}


\begin{frontmatter}

\title{{\bf \boldmath Measurement of the total and leptonic decay widths of the \jpsi resonance with an energy scan method at BESIII}}

\author{
\begin{small}
\begin{center}
M.~Ablikim$^{1}$, M.~N.~Achasov$^{11,b}$, P.~Adlarson$^{70}$, M.~Albrecht$^{4}$, R.~Aliberti$^{31}$, A.~Amoroso$^{69A,69C}$, M.~R.~An$^{35}$, Q.~An$^{66,53}$, X.~H.~Bai$^{61}$, Y.~Bai$^{52}$
% ...
\\
\vspace{0.2cm}
(BESIII Collaboration)\\
\vspace{0.2cm}
{\it
$^{1}$ Institute of High Energy Physics, Beijing 100049, People's Republic of China\\
% ...
$^{4}$ Bochum Ruhr-University, D-44780 Bochum, Germany\\
% ...
$^{11}$ G.I. Budker Institute of Nuclear Physics SB RAS (BINP), Novosibirsk 630090, Russia\\
% ...
$^{31}$ Johannes Gutenberg University of Mainz, Johann-Joachim-Becher-Weg 45, D-55099 Mainz, Germany\\
% ...
$^{35}$ Liaoning Normal University, Dalian 116029, People's Republic of China\\
% ...
$^{52}$ Southeast University, Nanjing 211100, People's Republic of China\\
$^{53}$ State Key Laboratory of Particle Detection and Electronics, Beijing 100049, Hefei 230026, People's Republic of China\\
% ...
$^{61}$ University of Jinan, Jinan 250022, People's Republic of China\\
% ...
$^{66}$ University of Science and Technology of China, Hefei 230026, People's Republic of China\\
% ...
$^{69}$ University of Turin and INFN, (A)University of Turin, I-10125, Turin, Italy; (B)University of Eastern Piedmont, I-15121, Alessandria, Italy; (C)INFN, I-10125, Turin, Italy\\
$^{70}$ Uppsala University, Box 516, SE-75120 Uppsala, Sweden\\
% ...
$^{b}$ Also at the Novosibirsk State University, Novosibirsk, 630090, Russia\\
% ...
}
\end{center}
\end{small}
}

\begin{abstract}

Using $e^+e^-$ annihilation data sets collected with the BESIII detector, we measure the cross sections of the processes \eetoee and \eetomumu at fifteen center-of-mass energy points in the vicinity of the \jpsi resonance. By a simultaneous fit to the measured, center-of-mass energy dependent cross sections of the two processes, the combined quantities $\EwMEwDTw$ and $\EwMMwDTw$ are determined to be ($0.346 \pm 0.009$) and ($0.335 \pm 0.006$) keV, respectively, where $\Ew$, $\Mw$, and $\Tw$ are the electronic, muonic, and total decay widths of the \jpsi resonance, respectively. Using the resultant $\EwMEwDTw$ and $\EwMMwDTw$, the ratio $\EwDMw$ is calculated to be $1.031 \pm 0.015$, which is consistent with the expectation of lepton universality within about two standard deviations. Assuming lepton universality and using the branching fraction of the \jpsi leptonic decay measured by BESIII in 2013, $\Tw$ and $\Lw$ are determined to be ($93.0 \pm 2.1$) and ($5.56 \pm 0.11$) keV, respectively, where $\Lw$ is the average leptonic decay width of the \jpsi resonance.
\ \\
\ \\
\text{Keywords: \jpsi, decay width, lepton universality, energy scan, BESIII}

\end{abstract}

\end{frontmatter}

\begin{multicols}{2}

\section{Introduction}
The total and electronic decay widths $\Tw$ and $\Ew$ of the \jpsi resonance, present in the Breit-Wigner formulae for all the decay modes of \jpsi produced in $e^{+}e^{-}$ collisions~\cite{PDG}, are among its most important parameters.
Theoretically, these decay widths, reflecting \jpsi internal interactions, are predicted by various potential models~\cite{potential models 0, potential models 1,potential models 2, the Cornell potential model, Richardson's potential model} and lattice quantum chromodynamics~\cite{lattice calculations}. 
Measurements of these decay widths and comparisons of the experimental results with different theoretical calculations can help us gain a better understanding of the underlying physics.

Furthermore, the ratio of the electronic to muonic decay widths of the \jpsi resonance, $\EwDMw$, can be used to test the lepton universality assumption~\cite{lepton universality}. 
Based on the assumption, the ratio is derived to be~\cite{formula for the leptonic decay widths of vector mesons}
\begin{equation}
\EwDMw = \frac{\beta_e(3-\beta_e^2)}{\beta_{\mu}(3-\beta_{\mu}^2)},
\end{equation}
\noindent with
\begin{equation}
\beta_l = \sqrt{1-(2m_l/M)^2}\text{\ \ \ \ \ }l=e,\mu,
\end{equation}
\noindent where $m_e$, $m_{\mu}$, and $M$ are the masses of electron, muon, and the \jpsi resonance, respectively. Taking the values of $m_e$, $m_{\mu}$, and $M$ from the Particle Data Group (PDG)~\cite{PDG}, $\EwDMw$ is calculated to be 1.00000814211(6), which has a deviation from 1 that is far less than the experimental precision at present. 
Thus, any observed, significant deviation of $\EwDMw$ from 1 will be a hint of physics beyond the Standard Model~\cite{charged lepton flavour violation}.

Since the discovery of the \jpsi resonance in 1974~\cite{the discovery of Jpsi 1,the discovery of Jpsi 2}, its decay widths have been measured by many experiments~\cite{Mark1,FRAG,FRAM,DASP,BES}.
The precision of the measurements has been improved significantly in the past two decades.
In 2004 and 2006, the \jpsi decay widths have been measured by studying the \jpsi samples produced in the initial state radiation (ISR) return process \eetomumugam collected at the $\Upsilon(4S)$ and $\psi(3770)$ peaks by BaBar~\cite{BaBar} and CLEO~\cite{CLEO}, respectively. In 2010, KEDR improved the measurement precision by performing an energy scan (ES) around the \jpsi peak and studying the \jpsi production in the processes \eetoee and \eetomumu~\cite{KEDR}. In 2018 and 2020, KEDR presented new results with the \jpsi production in the processes \eetoee and \eetohad~\cite{KEDR2}.

Operating in the $\tau$-charm energy region, the high luminosity of the BEPCII collider~\cite{BEPCII collider} and the excellent performance of the BESIII detector~\cite{BESIII detector} offer us a good opportunity for the precision measurements of the \jpsi decay widths. In 2016, BESIII measured the \jpsi decay widths by applying the ISR return technique to the data sample collected at the $\psi(3770)$ peak, and obtained a result with improved precision~\cite{BESIII}. 
In this Letter, we report a new precision measurement of the \jpsi decay widths with the ES method, confirming and complementing the above measurement.

Since the \jpsi resonance contributes to the vacuum polarization in the time-like region, the cross sections of the processes \eetoee and \eetomumu are functions of the \jpsi decay widths~\cite{Feynman,KEDRpsip}. 
Specifically, the cross section ($\sigma_{\rm 0}$) of each process with respect to the center-of-mass (CM) energy ($W_{\rm 0}$) can be written as
\begin{equation}
\sigma_{\rm 0}(W_{\rm 0}) = \sigma_{\rm 0}^{\rm C}(W_{\rm 0}) + \sigma_{\rm 0}^{\rm R}(W_{\rm 0}) + \sigma_{\rm 0}^{\rm I}(W_{\rm 0}) \label{Equation: ISR},
\end{equation}
where $\sigma_{\rm 0}^{\rm C}$, $\sigma_{\rm 0}^{\rm R}$ and $\sigma_{\rm 0}^{\rm I}$ are the continuum, resonance and interference terms, respectively. The formula still holds after considering the ISR effect, and we take $\sigma_{\rm 0}$, $\sigma_{\rm 0}^{\rm C}$, $\sigma_{\rm 0}^{\rm R}$ and $\sigma_{\rm 0}^{\rm I}$ here as the quantities with ISR considered. 
Unlike the term $\sigma_{\rm 0}^{\rm C}$, the terms $\sigma_{\rm 0}^{\rm R}$ and $\sigma_{\rm 0}^{\rm I}$ depend on the \jpsi decay widths, and their analytic forms are derived in Ref.~\cite{Analytic Forms for Cross Sections of Di-lepton Production from e+e- Collision around the Jpsi Resonance} using the structure function method~\cite{STRUCTUREFUNCTIONMETHOD1,STRUCTUREFUNCTIONMETHOD2}.

In Eq.~(\ref{Equation: ISR}), $\sigma_{\rm 0}^{\rm R}$ is the primary term related to the \jpsi decay widths, and its major subterm is proportional to $\EwMEwDTw$ and $\EwMMwDTw$ for the processes \eetoee and \eetomumuwos, respectively~\cite{Analytic Forms for Cross Sections of Di-lepton Production from e+e- Collision around the Jpsi Resonance}. Therefore, we can determine $\EwMEwDTw$ and $\EwMMwDTw$ by fitting to the measured, CM energy dependent cross sections of the two processes. Then, $\EwDMw$ can be evaluated as the ratio of $\EwMEwDTw$ to $\EwMMwDTw$. 
Combined with the branching fraction of the \jpsi leptonic decay measured by BESIII in 2013~\cite{BESIII --- Branch ratio of Jpsi to ll}, the total and leptonic \jpsi decay widths can be obtained from $\EwMEwDTw$ and $\EwMMwDTw$ as well.

\section{Experimental facilities and data sets}
\label{Section: Experimental Facilities and Data Sets}
The data used in this work were collected with the BESIII detector~\cite{BESIII detector}, which operates at the south crossing point of the BEPCII collider~\cite{BEPCII collider}. BEPCII is a double-ring $e^+e^-$ collider operating in the $\tau$-charm energy region (2.0-4.9 GeV) with its achieved peak luminosity of $10^{33}\text{ cm}^{-2}\text{s}^{-1}$ at the CM energy $\sqrt{s}=3.773$ GeV. The BESIII detector with a geometrical acceptance of 93\% of 4$\pi$, consists of the following main components: (1) a small-cell, helium-based main drift chamber (MDC) measuring the momenta of charged tracks in a 1 T magnetic field with a resolution of 0.5\% for 1 GeV/$c$ transverse momentum and the specific energy loss ($dE/dx$) with a resolution of 6\%; (2) a time-of-flight (TOF) system for particle identification composed of a barrel and two endcaps made of plastic scintillators; the time resolution is 80 ps in the barrel, and 110 ps in the endcaps; (3) an electromagnetic calorimeter (EMC) made of CsI(Tl) crystals arranged in a cylindrical shape (barrel) and two endcaps; for 1.0 GeV photons, the energy resolution is 2.5\% in the barrel and 5\% in the endcaps; (4) a superconducting solenoid magnet providing a nominal magnetic field of 1 T (0.9 T in 2012) parallel to the beam direction; (5) a muon chamber system made of resistive plate chambers with position resolution about 2 cm.

In addition, a beam energy measurement system (BEMS), located at the north crossing point of the BEPCII storage rings, is used to determine the BEPCII beam energies by measuring the energies of Compton back-scattered photons~\cite{BEMS}.

In 2012, an ES experiment was performed at fifteen CM energy points in the vicinity of the \jpsi resonance. The measured CM  energies and integrated luminosities are \  summarized \  in \  Table ~\ref{Table: Basic information of data.}. \  The \  CM \ \  energies \ \  are 

\end{multicols}
\begin{table}[!h]
\captionsetup{width=0.725\textwidth}
\caption{\small Basic information of the data samples at all individual CM energy points, including the proposed CM energies (Prop. $\sqrt{s}$) before the data taking, the CM energies measured by the BEMS and calibrated with the \jpsi mass from the PDG (Calib. BEMS $\sqrt{s}$), and the integrated luminosities (Int. $L$). For Calib. BEMS $\sqrt{s}$ and Int. $L$, the first uncertainties are statistical and the second ones are systematic.}
\centering
\begin{tabular*}{0.725\textwidth}{@{\extracolsep{\fill}}lcr}
\hline
Prop. $\sqrt{s}$ (MeV) & Calib. BEMS $\sqrt{s}$ (MeV) & Int. $L$ (pb$^{-1}$) \\
\hline
3050.0 & 3049.642$\pm$0.026$\pm$0.033 & 14.919$\pm$0.029$\pm$0.158 \\
3060.0 & 3058.693$\pm$0.028$\pm$0.033 & 15.060$\pm$0.029$\pm$0.158 \\
3083.0 & 3082.496$\pm$0.023$\pm$0.033 & 4.769$\pm$0.017$\pm$0.052 \\
3090.0 & 3088.854$\pm$0.022$\pm$0.033 & 15.558$\pm$0.030$\pm$0.162 \\
3093.0 & 3091.760$\pm$0.025$\pm$0.033 & 14.910$\pm$0.030$\pm$0.157 \\
3094.3 & 3094.697$\pm$0.084$\pm$0.033 & 2.143$\pm$0.011$\pm$0.023 \\
3095.2 & 3095.430$\pm$0.081$\pm$0.033 & 1.816$\pm$0.010$\pm$0.019 \\
3095.8 & 3095.826$\pm$0.075$\pm$0.033 & 2.135$\pm$0.011$\pm$0.023 \\
3096.9 & 3097.213$\pm$0.076$\pm$0.033 & 2.069$\pm$0.011$\pm$0.024 \\
3098.2 & 3098.340$\pm$0.075$\pm$0.033 & 2.203$\pm$0.011$\pm$0.023 \\
3099.0 & 3099.042$\pm$0.093$\pm$0.033 & 0.756$\pm$0.007$\pm$0.008 \\
3101.5 & 3101.359$\pm$0.106$\pm$0.033 & 1.612$\pm$0.010$\pm$0.018 \\
3105.5 & 3105.580$\pm$0.090$\pm$0.033 & 2.106$\pm$0.011$\pm$0.022 \\
3112.0 & 3112.051$\pm$0.093$\pm$0.033 & 1.720$\pm$0.010$\pm$0.019 \\
3120.0 & 3119.878$\pm$0.115$\pm$0.033 & 1.264$\pm$0.009$\pm$0.013 \\
\hline
\end{tabular*}
\label{Table: Basic information of data.}
\end{table}

\begin{multicols}{2}

\noindent measured by the BEMS and calibrated according to the \jpsi mass value given by the PDG~\cite{PDG}. As a consequence, the \jpsi mass can not be \noindent determined in this work. The CM energy calibration process fits to the \jpsi lineshapes in the $e^+e^-$ and $\mu^+\mu^-$ final states simultaneously to their preliminary measured, CM energy dependent cross sections. Adding the uncertainties of the \jpsi masses from the fit and from the PDG in quadrature gives a total calibration uncertainty of 0.033 MeV, which is comparable with the systematic uncertainty (0.043 MeV) of the calibration (via the inclusive hadronic decay mode) to the small \jpsi scan data used for the $\tau$ mass measurement~\cite{TAUMASS}. The corresponding integrated luminosities are measured offline with \eetodigam events~\cite{BESIII --- R scan --- luminosity}.

To determine the signal detection efficiencies, Monte Carlo (MC) simulated events of the processes \eetoee and \eetomumu in the polar angle ranges of 34$^{\circ}$-146$^{\circ}$ and 0$^{\circ}$-180$^{\circ}$, respectively, incorporating the ISR and final state radiation~(FSR) effects, are simulated with a revised version of the {\sc Babayaga-3.5}~\cite{BABAYAGA} generator, which is modified by the authors to explicitly involve the \jpsi resonance in the vacuum polarization. In addition, MC events of the processes \eetohad and \eetotwogam ($e^+e^-e^+e^-$, $e^+e^-\mu^+\mu^-$, etc.) are generated for background studies with the {\sc ConExc}~\cite{CEXC} and {\sc BesTwogam}~\cite{BESTWOGAMMA} generators, respectively. When generating these events, the calibrated BEMS CM energies are used.

  A {\sc geant4}~\cite{geant4} based MC simulation program including the geometric
  description and response of the detector is used to simulate the interaction of final state particles in the detector. Both the experimetal data and
  simulated MC are reconstructed and analysed under the
 {\sc gaudi}~\cite{GAUDI} based offline software system.

\section{Event selection}
\label{Section: Event Selection}
The signal candidates of \eetoee and \eetomumu events are required to have two oppositely charged tracks in the MDC. 
Each charged track has to fulfill the following requirements: 
it must originate from the interaction region of $|V_{\rm r}|<1$ cm and $|V_{\rm z}|<10$ cm, where $|V_{\rm r}|$ and $|V_{\rm z}|$ are its closest approach relative to the collision point in the x-y plane and along the z axis (taken as the axis of the MDC), respectively; 
it must hit the detector in the barrel region of $|\csth|<0.8$, where $\theta$ is the polar angle of the reconstructed momentum vector with respect to the z axis.

For \eetoee candidates, the next two criteria are applied further to each of the selected tracks: 
its momentum ($P$) is required to be larger than 0.7 times the beam energy ($E_{\rm beam}$), 
and its energy deposited in the EMC ($E$) has to be larger than 0.6 times the momentum.

For \eetomumu candidates, the following conditions are required in addition: 
for each of the selected tracks, $P$ must be larger than 0.8$E_{\rm beam}$, $E$ has to be larger than 25 MeV and less than 0.25$P$, and valid timing information is required to be left in the TOF; 
at the event level, no neutral showers with a deposited energy above 25 MeV are allowed in the EMC, 
and the difference of the flight time of the two charged tracks ($\Delta t_{\rm TOF}^{\mu}$) obtained with the TOF has to be less than $1.5$ ns to suppress cosmic rays.

\begin{figure}[!h]
\centering
\subfigure{\includegraphics[width=0.325\textwidth]{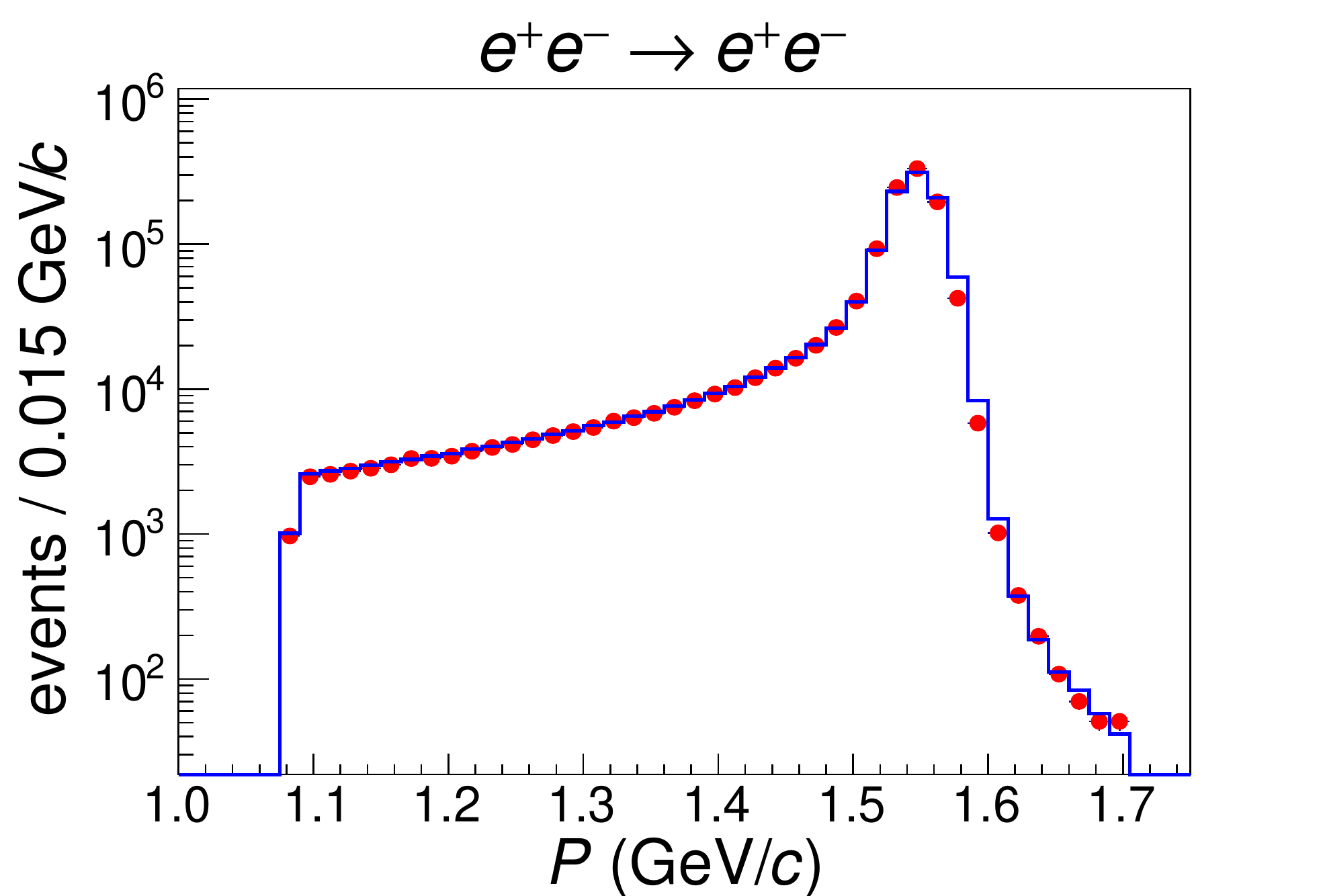}}
\subfigure{\includegraphics[width=0.325\textwidth]{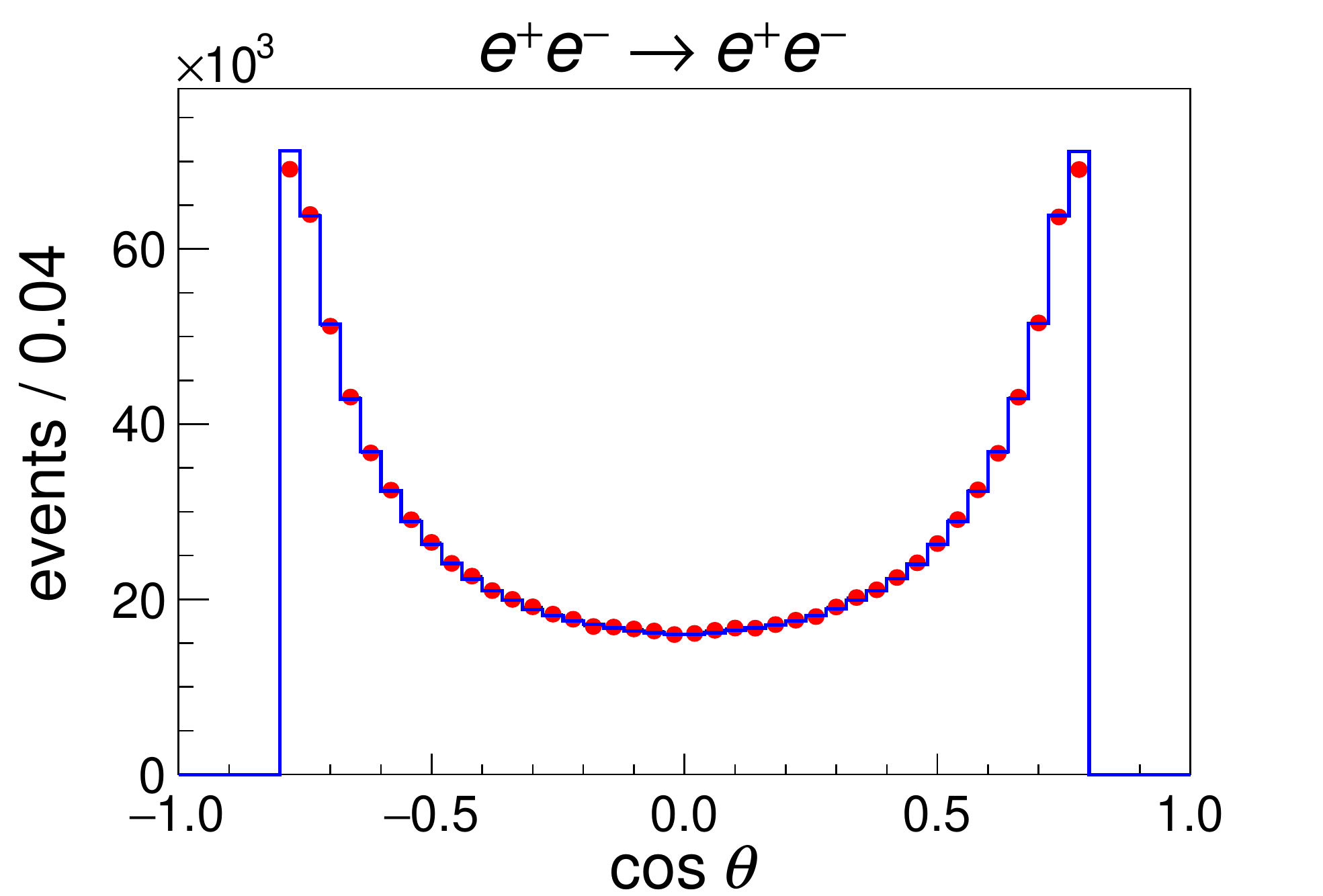}}
\subfigure{\includegraphics[width=0.325\textwidth]{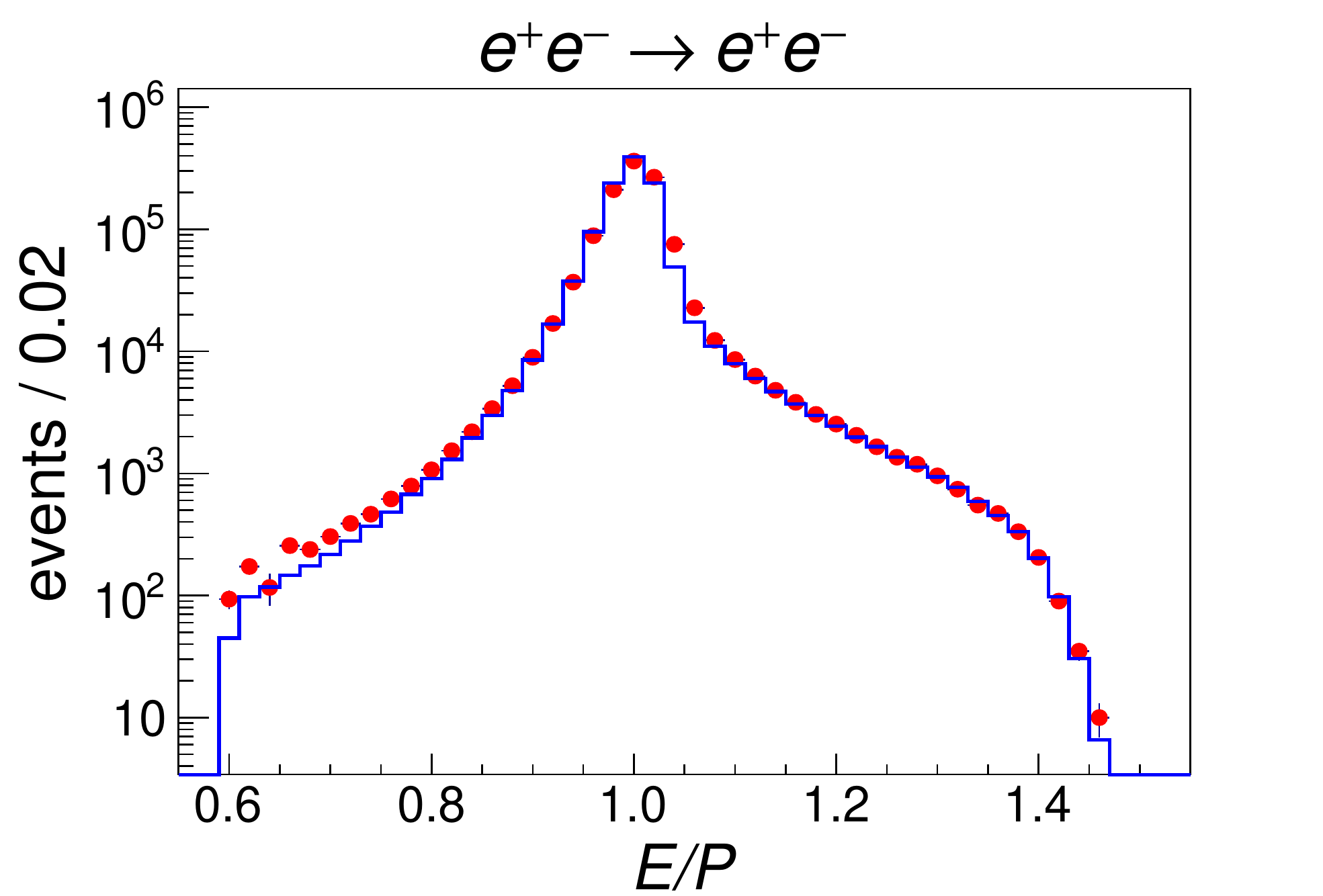}}
\subfigure{\includegraphics[width=0.325\textwidth]{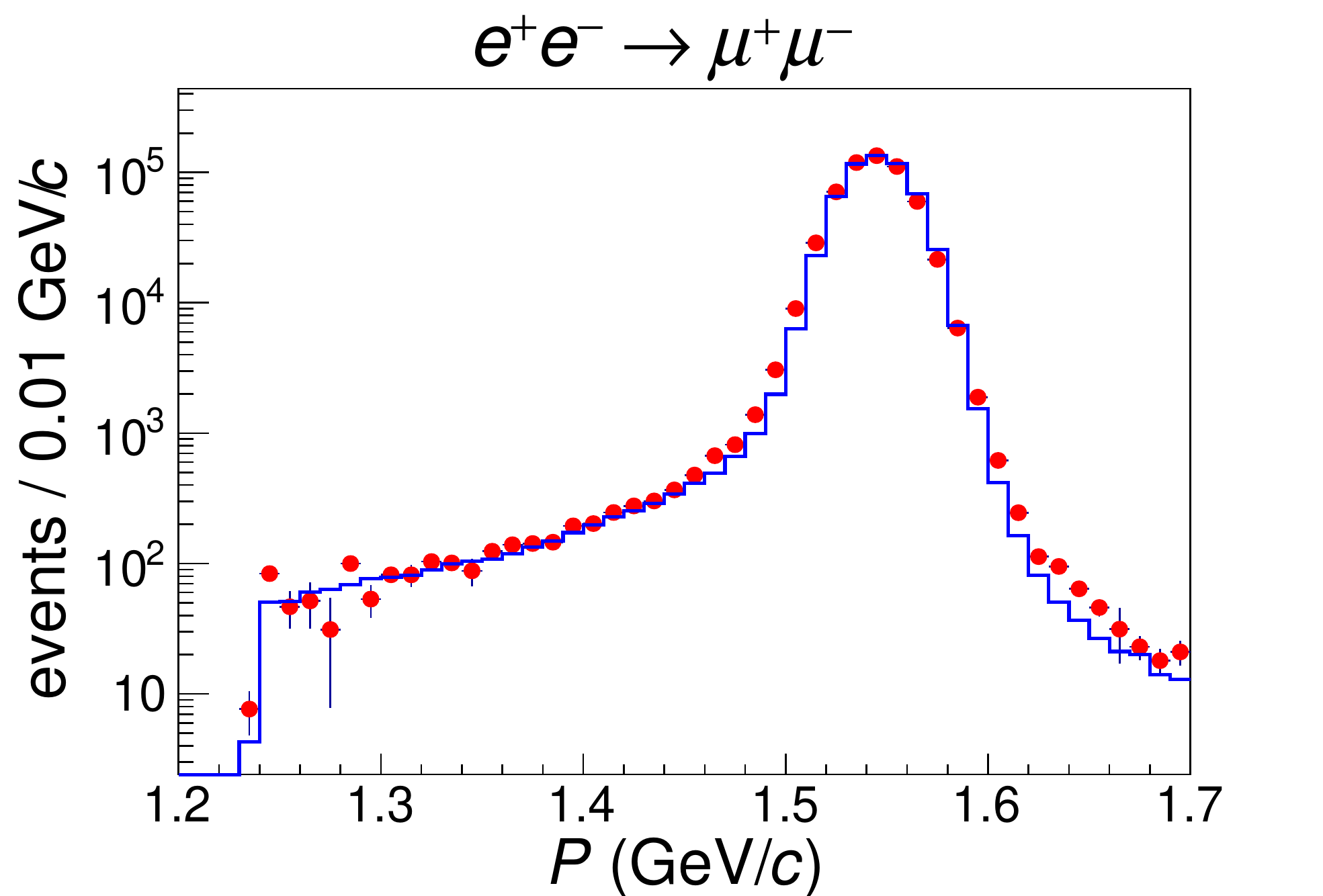}}
\subfigure{\includegraphics[width=0.325\textwidth]{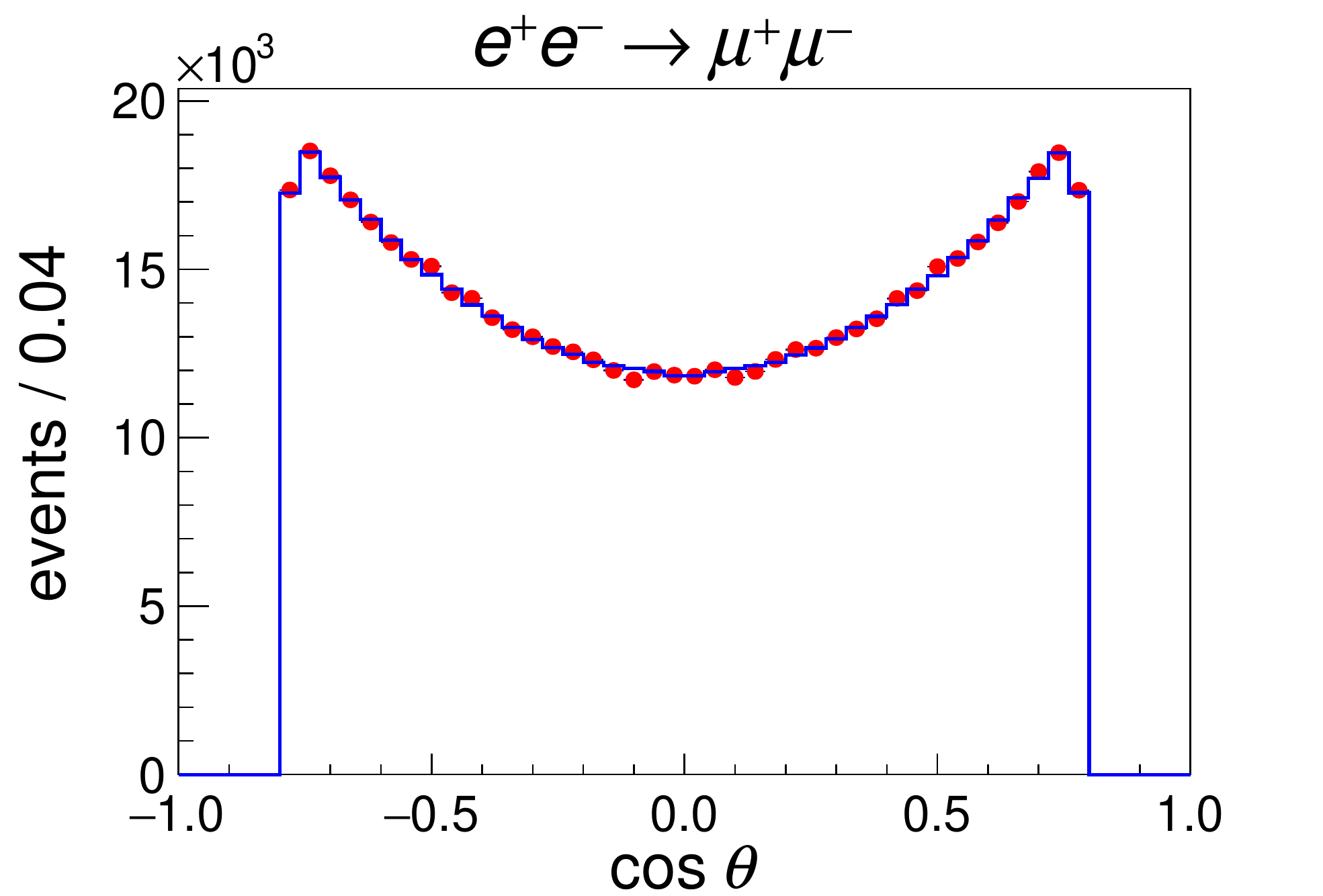}}
\subfigure{\includegraphics[width=0.325\textwidth]{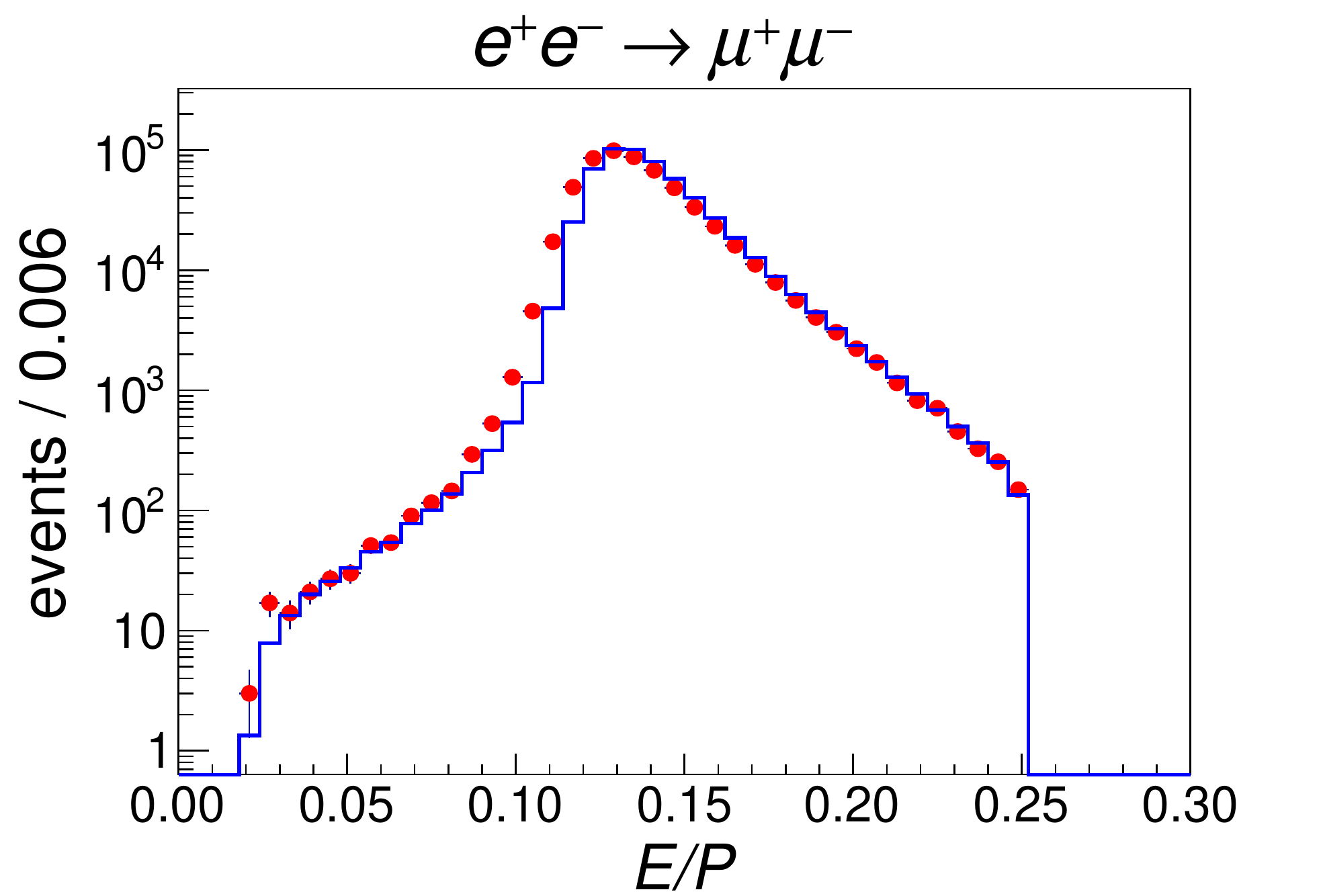}}
\subfigure{\includegraphics[width=0.325\textwidth]{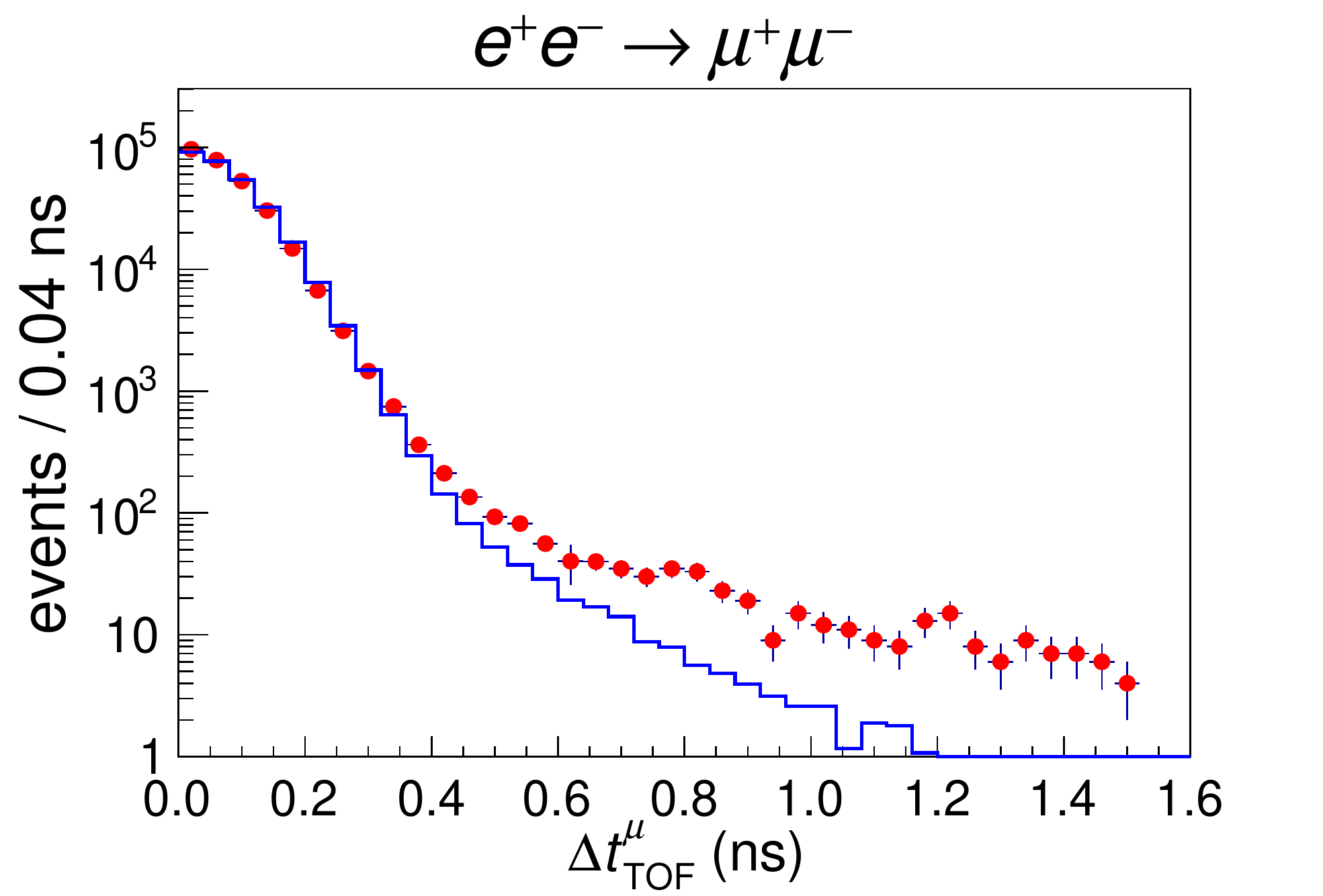}}
\caption{\small Comparison between data and MC simulation of the distributions of the variables used in the selection criteria for the processes \eetoee and \eetomumu at $\sqrt{s}=3096.9$ MeV. For all the plots, dots with error bars show the background-subtracted data (the background level is evaluated with the MC simulation) and the histograms denote the signal MC. The small discrepancy in the last plot is due to the imperfection of the MC simulation, and it has a negligible effect on the cross section measurement of $e^+e^- \to \mu^+\mu^-$.}
\label{Figure: Comparisons between data and MC of the selection condition related distributions of signal candidates of eetoee and eetomumu at sqrts=3096.9 MeV after event selection.}
\end{figure}

Figure~\ref{Figure: Comparisons between data and MC of the selection condition related distributions of signal candidates of eetoee and eetomumu at sqrts=3096.9 MeV after event selection.} shows the comparison between data and MC simulation of variables used in the event selection. 
The data shown in the figure are those of surviving candidate events subtracting the residual background estimated with the MC simulation. 
In general, the MC simulation provides a good description of the data although minor discrepancies between them are visible. 
The effects of these discrepancies are taken into account in the systematic uncertainty estimation (see Section~\ref{Subsection: Systematic uncertainties} for details).

For the process \eetoee ($\mu^+\mu^-$), the efficiencies obtained from the signal MC samples are about 70\% (80\%), and the background levels esitimated with the MC simulation are less than 0.05\% (0.5\%).
Closer examination with a generic event type analysis tool, {\sc TopoAna}~\cite{TopoAna}, shows that the backgrounds mainly arise from events with $\pi^+\pi^-$, $K^+K^-$ or $e^+e^-e^+e^-$ ($e^+e^-\mu^+\mu^-$) final states.

\section{Cross section measurement}
\label{Section: Cross Section Measurement}
\subsection{Nominal results with statistical uncertainties}
Usually, the cross section $\sigma$ is determined from
\begin{equation}
\sigma = \frac{N_{\rm sig} - N_{\rm bkgs}}{L \cdot \epsilon_{\rm trg} \cdot \epsilon_{\rm recsel}} \cdot f,
\label{Equation: nominal cross section formula}
\end{equation}
where $N_{\rm sig}$ is the number of signal events selected from data, $N_{\rm bkgs}$ is the number of the residual background events, $L$ is the integrated luminosity, $\epsilon_{\rm trg}$ is the trigger efficiency, $\epsilon_{\rm recsel}$ is the reconstruction-selection efficiency and $f$ is a reconstruction efficiency correction factor.

The trigger efficiency is taken as 100\% in this
work~\cite{TRGEFF}. The correction factor $f$ is related to the imperfection of
the detector simulation. In practice, the reconstruction efficiencies (including the tracking efficiency in the MDC and the cluster reconstruction efficiency in the EMC) from MC simulation deviate those from data. We \  study \  the \  corresponding \  reconstruction 

\end{multicols}

\begin{multicols}{2}

\noindent efficiencies for leptons in different $\cos\theta$ bins.
To compensate the deviation of the reconstruction-selection efficiency, the correction factor $f$ is introduced as
\begin{align}
f & = N^{\rm MC}_{\rm obs} \Bigg/ \Bigg(\sum \limits_m \sum \limits_n N^{\rm MC}_{\rm obs}(m,n) \cdot \frac{\epsilon_{\rm trk}^{\rm data}(m)}{\epsilon_{\rm trk}^{\rm MC}(m)} \nonumber \\
& \cdot \frac{\epsilon_{\rm trk}^{\rm data}(n)}{\epsilon_{\rm trk}^{\rm MC}(n)} \cdot \frac{\epsilon_{\rm clst}^{\rm data}(m)}{\epsilon_{\rm clst}^{\rm MC}(m)} \cdot \frac{\epsilon_{\rm clst}^{\rm data}(n)}{\epsilon_{\rm clst}^{\rm MC}(n)} \Bigg).
\label{Equation: Efficiency correction factor}
\end{align}

  \noindent Here, $N^{\rm MC}_{\rm obs}$ stands for the number of
surviving events of signal MC samples, $m$ ($n$) for the $m^{\rm th}$ ($n^{\rm
    th}$) $\cos\theta$ bin of positively (negatively) charged
  leptons, $\epsilon_{\rm trk}^{\rm data}$ and $\epsilon_{\rm
    trk}^{\rm MC}$ ($\epsilon_{\rm clst}^{\rm data}$ and
  $\epsilon_{\rm clst}^{\rm MC}$) for the MDC tracking efficiency (EMC
  cluster reconstruction efficiency) of leptons from data and MC simulation,
  respectively.

The measured cross sections and related input quantities at all individual CM energy points of the processes \eetoee and \eetomumu are summarized in Tables \ref{Table: Measured cross sections and related input quantities of the process eetoee at all individual CM energy points.} and \ref{Table: Measured cross sections and related input quantities of the process eetomumu at all individual CM energy points.}, respectively.

\subsection{Systematic uncertainties}
\label{Subsection: Systematic uncertainties}
The systematic uncertainties of the measured cross sections arise mainly from the integrated luminosities, trigger efficiencies, CM energies, reconstruction and selection efficiencies, efficiency correction factors and residual backgrounds.

The uncertainties due to the integrated luminosities are estimated to be less than 1.40\% (1.26\% at $\sqrt{s}=3096.9$ MeV)~\cite{BESIII --- R scan --- luminosity}, while those resulting from trigger efficiencies are evaluated as 0.10\%~\cite{TRGEFF}.

\end{multicols}
\begin{table}[!h]
\captionsetup{width=0.95\textwidth}
\caption{Measured cross sections and related input quantities of the process \eetoee at all individual CM energy points. 
The uncertainties of the input quantities are statistical, while the first and second uncertainties of cross sections are statistical and systematic, respectively. 
Notably, the cross sections here are confined in the polar angle range of 34$^{\circ}$--146$^{\circ}$.}
\centering
\begin{tabular*}{0.95\textwidth}{@{\extracolsep{\fill}}lccccr}
\hline
Prop. $\sqrt{s}$ (MeV) & $N_{\rm sig}$ & $N_{\rm bkgs}$ & $\epsilon_{\rm recsel}$ & $f$ & $\sigma$ (nb) \\
\hline
3.0500 & $2274639\pm1508$ & $164\pm13$ & $0.6909\pm0.0002$ & $0.9989\pm0.0002$ & $220.4\pm0.2\pm2.5$ \\
3.0600 & $2286953\pm1512$ & $145\pm12$ & $0.6910\pm0.0002$ & $0.9989\pm0.0002$ & $219.5\pm0.2\pm2.5$ \\
3.0830 & $710011\pm843$ & $57\pm8$ & $0.6913\pm0.0002$ & $0.9989\pm0.0002$ & $215.1\pm0.3\pm2.6$ \\
3.0900 & $2341309\pm1530$ & $182\pm13$ & $0.6918\pm0.0002$ & $0.9989\pm0.0002$ & $217.3\pm0.2\pm2.4$ \\
3.0930 & $2240003\pm1497$ & $181\pm13$ & $0.6918\pm0.0002$ & $0.9989\pm0.0002$ & $216.9\pm0.2\pm2.5$ \\
3.0943 & $345449\pm588$ & $34\pm6$ & $0.6989\pm0.0002$ & $0.9990\pm0.0002$ & $230.4\pm0.4\pm2.9$ \\
3.0952 & $335948\pm580$ & $52\pm7$ & $0.7180\pm0.0002$ & $0.9991\pm0.0002$ & $257.3\pm0.4\pm3.4$ \\
3.0958 & $487256\pm698$ & $61\pm8$ & $0.7329\pm0.0002$ & $0.9992\pm0.0002$ & $311.2\pm0.4\pm6.9$ \\
3.0969 & $577995\pm760$ & $174\pm13$ & $0.7580\pm0.0002$ & $0.9994\pm0.0002$ & $368.2\pm0.5\pm5.2$ \\
3.0982 & $443694\pm666$ & $104\pm10$ & $0.7286\pm0.0002$ & $0.9992\pm0.0002$ & $276.2\pm0.4\pm4.2$ \\
3.0990 & $127932\pm358$ & $19\pm4$ & $0.7082\pm0.0002$ & $0.9991\pm0.0002$ & $238.6\pm0.7\pm4.3$ \\
3.1015 & $242275\pm492$ & $26\pm5$ & $0.6958\pm0.0002$ & $0.9990\pm0.0002$ & $215.7\pm0.4\pm2.8$ \\
3.1055 & $313080\pm560$ & $32\pm6$ & $0.6929\pm0.0002$ & $0.9990\pm0.0002$ & $214.3\pm0.4\pm2.7$ \\
3.1120 & $251731\pm502$ & $21\pm5$ & $0.6919\pm0.0002$ & $0.9990\pm0.0002$ & $211.3\pm0.4\pm2.8$ \\
3.1200 & $185572\pm431$ & $13\pm4$ & $0.6913\pm0.0002$ & $0.9990\pm0.0002$ & $212.1\pm0.5\pm2.8$ \\
\hline
\end{tabular*}
\label{Table: Measured cross sections and related input quantities of the process eetoee at all individual CM energy points.}
\end{table}

\begin{table}[!h]
\captionsetup{width=0.96\textwidth}
\caption{Measured cross sections and related input quantities of the process \eetomumu at all individual CM energy points. 
The uncertainties of the input quantities are statistical, while the first and second uncertainties of cross sections are statistical and systematic, respectively. 
Here, the cross sections are defined in the polar angle range of 0$^{\circ}$--180$^{\circ}$.}
\centering
\begin{tabular*}{0.96\textwidth}{@{\extracolsep{\fill}}lccccr}
\hline
Prop. $\sqrt{s}$ (MeV) & $N_{\rm sig}$ & $N_{\rm bkgs}$ & $\epsilon_{\rm recsel}$ & $f$ & $\sigma$ (nb) \\
\hline
3.0500 & $85997\pm293$ & $313\pm18$ & $0.5359\pm0.0002$ & $1.0094\pm0.0003$ & $10.82\pm0.04\pm0.12$ \\
3.0600 & $85639\pm293$ & $316\pm18$ & $0.5360\pm0.0002$ & $1.0094\pm0.0003$ & $10.67\pm0.04\pm0.12$ \\
3.0830 & $24413\pm156$ & $93\pm10$ & $0.5286\pm0.0002$ & $1.0094\pm0.0003$ & $9.74\pm0.06\pm0.12$ \\
3.0900 & $72967\pm270$ & $344\pm19$ & $0.5202\pm0.0002$ & $1.0094\pm0.0003$ & $9.06\pm0.03\pm0.11$ \\
3.0930 & $60436\pm246$ & $336\pm18$ & $0.5056\pm0.0002$ & $1.0094\pm0.0003$ & $8.05\pm0.03\pm0.10$ \\
3.0943 & $17432\pm132$ & $82\pm9$ & $0.5781\pm0.0002$ & $1.0093\pm0.0003$ & $14.13\pm0.11\pm0.35$ \\
3.0952 & $53866\pm232$ & $176\pm13$ & $0.6299\pm0.0002$ & $1.0093\pm0.0003$ & $47.36\pm0.20\pm0.66$ \\
3.0958 & $158653\pm398$ & $318\pm18$ & $0.6366\pm0.0002$ & $1.0093\pm0.0003$ & $117.59\pm0.29\pm1.51$ \\
3.0969 & $287418\pm536$ & $728\pm27$ & $0.6418\pm0.0002$ & $1.0093\pm0.0003$ & $217.88\pm0.41\pm2.81$ \\
3.0982 & $158804\pm399$ & $294\pm17$ & $0.6369\pm0.0002$ & $1.0093\pm0.0003$ & $114.03\pm0.29\pm1.39$ \\
3.0990 & $27505\pm166$ & $58\pm8$ & $0.6216\pm0.0002$ & $1.0093\pm0.0003$ & $58.92\pm0.35\pm0.85$ \\
3.1015 & $23087\pm152$ & $119\pm11$ & $0.5981\pm0.0002$ & $1.0093\pm0.0003$ & $24.04\pm0.16\pm0.36$ \\
3.1055 & $21595\pm147$ & $139\pm12$ & $0.5786\pm0.0002$ & $1.0093\pm0.0003$ & $17.77\pm0.12\pm0.26$ \\
3.1120 & $14357\pm120$ & $69\pm8$ & $0.5674\pm0.0002$ & $1.0093\pm0.0003$ & $14.78\pm0.12\pm0.21$ \\
3.1200 & $9352\pm97$ & $53\pm7$ & $0.5570\pm0.0002$ & $1.0093\pm0.0003$ & $13.33\pm0.14\pm0.19$ \\
\hline
\end{tabular*}
\label{Table: Measured cross sections and related input quantities of the process eetomumu at all individual CM energy points.}
\end{table}

\begin{table}[!h]
\captionsetup{width=0.725\textwidth}
\caption{\small Relative systematic uncertainties of the measured cross sections of the processes \eetoee and \eetomumu at $\sqrt{s}=3096.9$ MeV.}
\centering
\begin{tabular*}{0.725\textwidth}{@{\extracolsep{\fill}}lcc}
\hline
\multirow{2}{*}{Source} & \multicolumn{2}{c}{Uncertainty (\%)} \\
& \eetoee & \eetomumu \\
\hline
Integrated luminosity measurement & 1.26 & 1.26 \\
Trigger efficiency determination & 0.10 & 0.10 \\
CM energy measurement & 0.17 & 0.11 \\
$\csth$ requirement & 0.55 & 0.01 \\
$P$ requirement & 0.13 & 0.02 \\
$E/P$ requirement & 0.05 & 0.01 \\
$\Delta t_{\rm TOF}^{\mu}$ requirement & --- & 0.05 \\
Efficiency correction factor $f$ & 0.10 & 0.10 \\
Residual backgrounds & 0.03 & 0.25 \\
\hline
Total & 1.40 & 1.30 \\
\hline
\end{tabular*}
\label{Table: Systematic uncertainties of the measured cross sections of eetoee and eetomumu at sqrts=3096.9 MeV}
\end{table}

\begin{multicols}{2}

To estimate the uncertainties due to the CM energies, two additional sets of MC samples are generated by increasing or decreasing the CM energies by one standard deviation with respect to their nominal values.
The largest changes of the efficiencies with respect to their nominal values are taken as the uncertainties.

The uncertainties associated with the momentum requirement for the process \eetoee are estimated by changing the selection criteria from $P>0.7E_{\rm beam}$ to $P>0.6 E_{\rm beam}$. The resultant changes in the calculated cross sections are taken as the uncertainties. The uncertainties related to other requirements are estimated with the similar method. Specifically, for the selection of \eetoee events, the analysis is carried out with the alternative criteria of $|\cos\theta|<0.7$ and $E/P<0.7$, individually, while for \eetomumuwos, the analysis is repeated with the alternative criteria of  $P>0.9 E_{\rm beam}$, $|\csth|<0.7$, $E/P<0.35$ \ \  and \ \  $\Delta t_{\rm TOF}^{\mu}<2.5$ ns, individually.

As shown in Tables \ref{Table: Measured cross sections and related input quantities of the process eetoee at all individual CM energy points.} and \ref{Table: Measured cross sections and related input quantities of the process eetomumu at all individual CM energy points.}, the statistical uncertainties of the efficiency correction factors are 0.02\% and 0.03\% for the processes \eetoee and \eetomumuwos, respectively, which are determined from the statistics of the samples used to study the reconstruction efficiencies.
On the other hand, detailed studies show that the purities of the control samples for the electron tracking, electron clustering, muon tracking, and muon clustering efficiencies in data are about 99.99\%, 99.81\%, 98.45\%, and 99.52\%, respectively. 
Considering other factors, such as the background contaminations, the uncertainties resulting from the efficiency correction factors can be roughly and conservatively estimated to be 0.10\% for both the \eetoee and \eetomumu processes.

The numbers of residual background events, estimated with the MC simulation, are subtracted from the numbers of surviving events in the calculation of the cross sections, and hence the uncertainties of background levels need be taken into account. Since the uncertainties of the cross sections for some dominant background channels (for example, $e^+ e^- \to K^+ K^-$) set in the generator are as large as 100\%, we therefore take the background levels themselves as the related uncertainties. As a result, the uncertainties for the processes \eetoee and \eetomumu at $\sqrt{s}=3096.9$ MeV are 0.03\% and 0.25\%, respectively.

Table~\ref{Table: Systematic uncertainties of the measured cross sections of eetoee and eetomumu at sqrts=3096.9 MeV} shows a summary of the systematic uncertainties of the measured cross sections of the processes \eetoee and \eetomumu at $\sqrt{s}=3096.9$ MeV. The total systematic uncertainties, 1.40\% and 1.29\% for the two processes individually, are the square root of the quadratic sum of the individual uncertainties and dominated by those associated with the integrated luminosities.
The systematic uncertainties of the measured cross sections at other CM energy points are estimated with the same method, and they are summarized in Tables \ref{Table: Measured cross sections and related input quantities of the process eetoee at all individual CM energy points.} and \ref{Table: Measured cross sections and related input quantities of the process eetomumu at all individual CM energy points.} together with the statistical uncertainties.

\subsection{Correlation analysis}
To consider the correlations between the measured cross sections of the same process at different CM energy points, the corresponding covariance matrices are estimated. To estimate such a covariance matrix, contributions from all related uncertainty sources are analysed and estimated according to their nature and the method of uncertainty propagation. To get an impression of the strength of these correlations, the correlation coefficient matrices of the measured cross sections of the processes \eetoee and \eetomumu are shown in Fig.~\ref{Figure: Correlation coefficient matrix of the measured cross sections of eetoee and eetomumu.}. We find that the correlations are strong and can not be neglected.

In the covariance matrix analysis above, the corresponding covariance matrix of the measured luminosities at different CM energy points is estimated in advance with the similar method. This matrix is required when estimating the covariance matrices of the measured cross sections and constructing the global $\chi^2$ function for the simultaneous fit of the processes \eetoee and \eetomumu (see Section~\ref{Subsection: Global Chisquare Function}).

\end{multicols}
\begin{figurehere}
\centering
\includegraphics[width=0.375\textwidth]{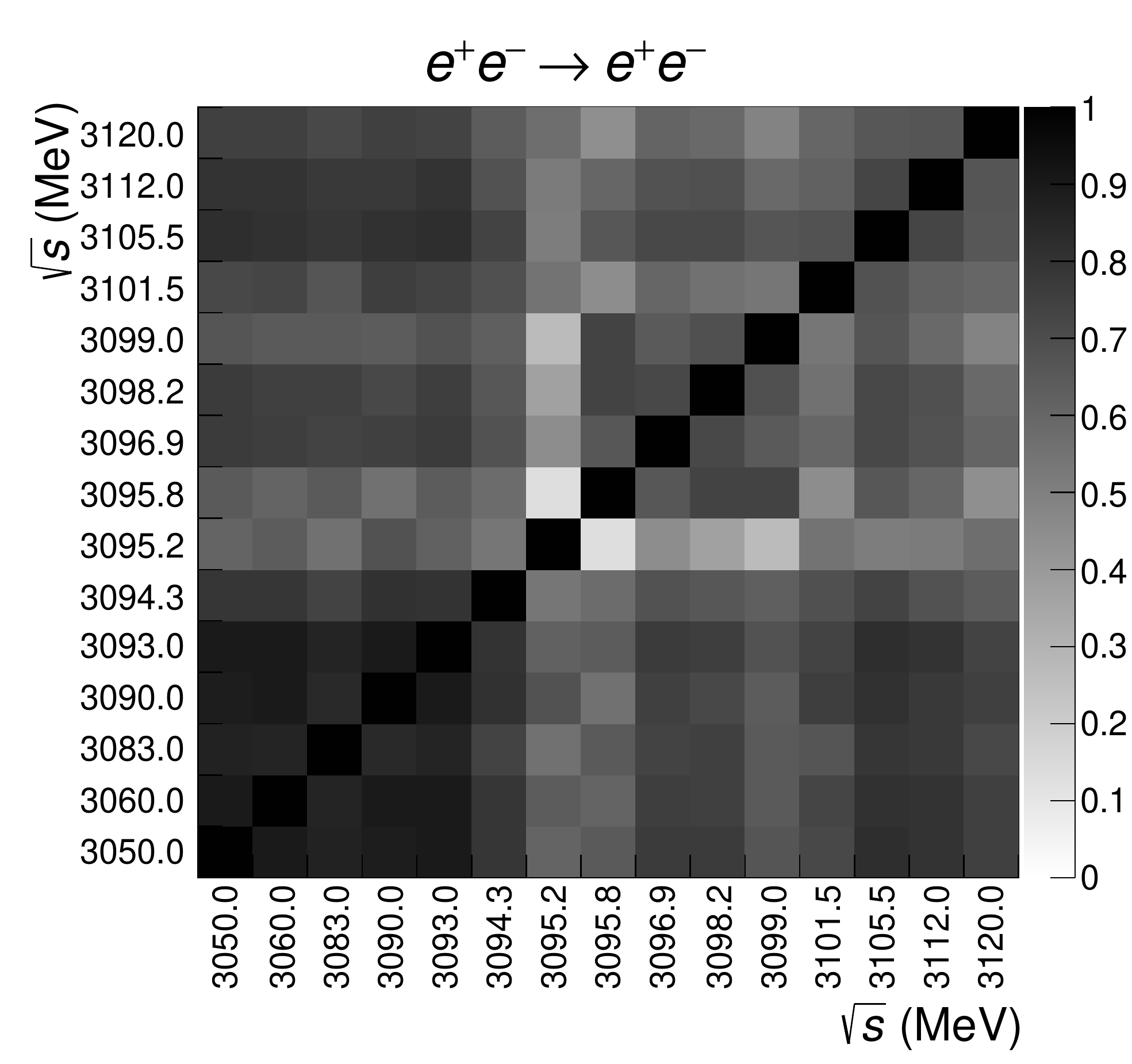}
\includegraphics[width=0.375\textwidth]{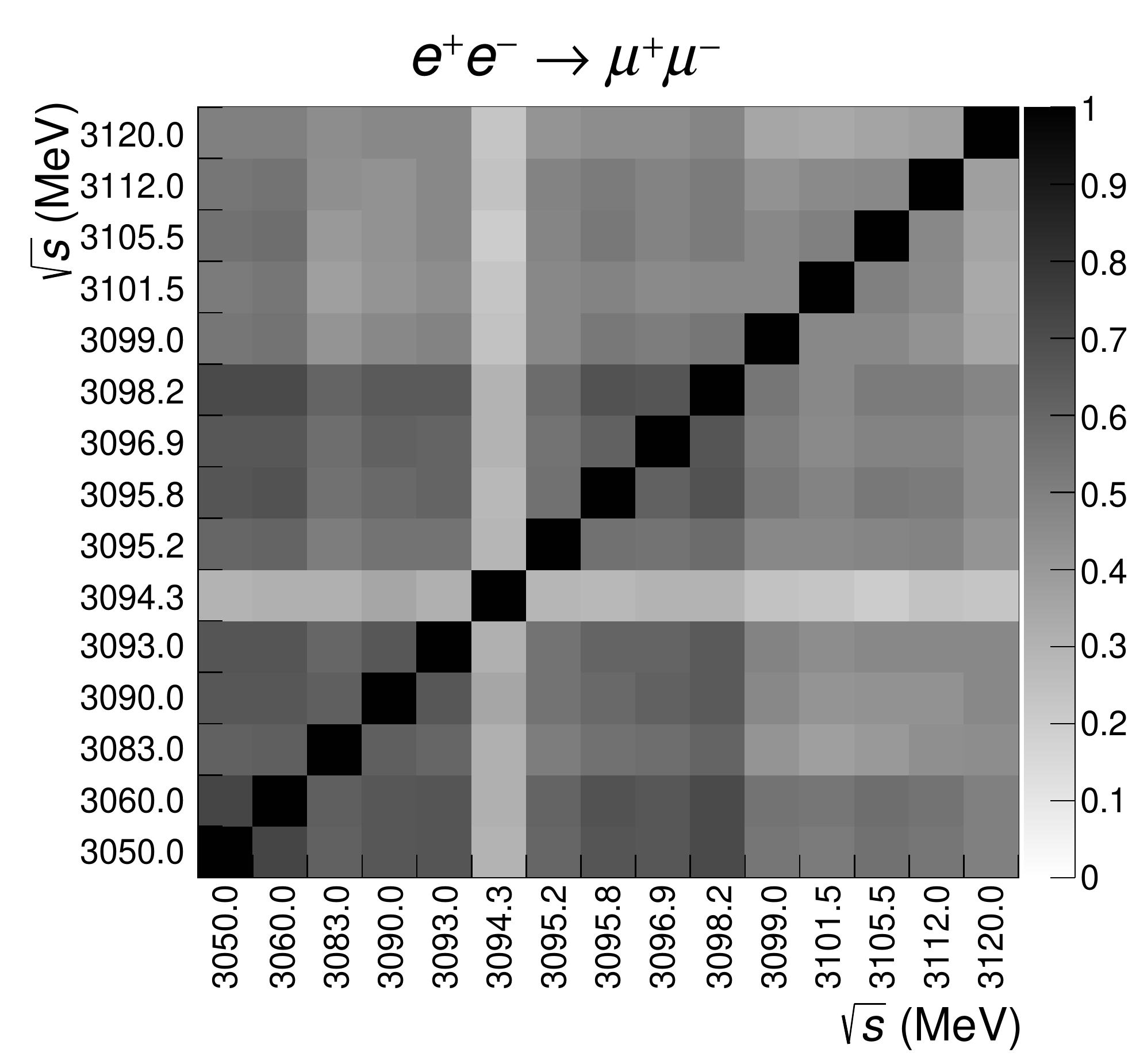}
\caption{\small Correlation coefficient matrices of the measured cross sections of the processes \eetoee and \eetomumuwos.}
\label{Figure: Correlation coefficient matrix of the measured cross sections of eetoee and eetomumu.}
\end{figurehere}

\begin{multicols}{2}

\section{Decay width determination}
\label{Section: Decay Width Extraction}
\subsection{Energy spread and final state radiation}
\label{Subsection: Energy Spread and Final State Radiation}
To determine the \jpsi decay widths, a simultaneous fit to the measured,
CM energy dependent cross sections of the processes \eetoee and
\eetomumu is required. In the theoretical formulae used in the fit,
the effects of the beam energy spread and FSR are taken into account as well.

By assuming the CM energy spread follows a Gaussian distribution, the theoretical cross section is
\begin{equation}
\sigma(W) = \int \sigma_{\rm 0}(W_{\rm 0}) \left(\frac{1}{\sqrt{2\pi}S_{\rm W}}\text{e}^{-\frac{(W_{\rm 0}-W)^2}{2S_{\rm W}^2}}\right) dW_{\rm 0}.
\label{Equation: ES}
\end{equation}
\noindent Here, $W$ ($=\sqrt{s}$) and $S_{\rm W}$ are the mean and standard deviation of the CM energy distribution, respectively. According to the formula and the expression of $\sigma_{\rm 0}$ in Eq. (\ref{Equation: ISR}), $\sigma$ can also be divided into three terms: the continuum term ($\sigma^{\rm C}$), the resonance term ($\sigma^{\rm R}$) and interference term ($\sigma^{\rm I}$). In practice, $\sigma^{\rm C}$ is evaluated with the {\sc Babayaga-3.5} generator~\cite{BABAYAGA} with the effects of \jpsi and FSR switched off, while $\sigma^{\rm R} + \sigma^{\rm I}$ is calculated using the analytic formulae for $\sigma_{\rm 0}^{\rm R} + \sigma_{\rm 0}^{\rm I}$ in Ref.~\cite{Analytic Forms for Cross Sections of Di-lepton Production from e+e- Collision around the Jpsi Resonance}.

In Eq. (\ref{Equation: ISR}), only the ISR effect is involved in $\sigma_{\rm 0}$. To take into account the FSR effect, we introduce a correction factor $R^{\rm FSR}$ into the theoretical cross section:
\begin{equation}
\sigma_{\rm 0}^{\rm theor} (W) = \sigma (W) \cdot R^{\rm FSR}(W).
\label{Equation: FSR}
\end{equation}
In practice, $R^{\rm FSR}$ is obtained with the {\sc Babayaga-3.5} generator~\cite{BABAYAGA} as the ratio of the calculated cross sections with and without the FSR effect. For example, at $\sqrt{s}=3096.9$ MeV, $R^{\rm FSR}$ is 0.980 and 0.998 for the processes \eetoee and \eetomumuwos, respectively.

Due to the high-order corrections related to the FSR effect, the cross sections of the processes \eetoee and \eetomumu are calculated by the {\sc Babayaga-3.5} generator with the uncertainties of 0.5\% and 1.0\%, respectively~\cite{BABAYAGA}.
Thus, systematic deviations of $R^{\rm FSR}(W)$ from their truth values probably appear in the vicinity of the \jpsi resonance.
To take the possible deviations into consideration, we implement one free scaling parameter in the theoretical cross section formula of each process for the simultaneous fit,
\begin{equation}
\sigma^{\rm theor} (W) = \sigma_{\rm 0}^{\rm theor} (W) \cdot F.
\label{Equation: FSR_correction}
\end{equation}
Specifically, the free scaling parameters for the processes \eetoee and \eetomumu are referred to as $F_{ee}$ and $F_{\mu\mu}$, respectively.
\subsection{Global $\chi^2$ function}
\label{Subsection: Global Chisquare Function}
To take into account the correlations between the measured cross sections of different processes and/or at different CM energy points, as well as the uncertainties of the CM energies, we construct a global $\chi^2$ function for the simultaneous fit according to the standard covariance matrix method as
\begin{equation}
\chi^2 = \Delta\sigma^{\rm T} \cdot V^{-1} \cdot \Delta\sigma,
\end{equation}
where
\end{multicols}
\begin{equation}
\Delta\sigma(i) =
\left\{
\begin{array}{ll}
\displaystyle \sigma_{ee}^{\rm exper}(i) - \sigma_{ee}^{\rm theor}(i) & \displaystyle i=1,2 \cdots 14,15 \\
\displaystyle \sigma_{\rm \mu\mu}^{\rm exper}(i-15) - \sigma_{\rm \mu\mu}^{\rm theor}(i-15) & \displaystyle i=16,17 \cdots 29,30 \\
\end{array}
\right.
\end{equation}
and
\begin{equation}
V(i,j) =
\left\{
\begin{array}{ll}
\displaystyle V_{ee}(i,j) + \delta(i,j) \, \left(\frac{\partial \sigma_{ee}^{\rm theor}}{\partial W}(i) \, \Delta W(i)\right)^2 & \displaystyle \textcircled{\footnotesize{1}} \\
\displaystyle \frac{\sigma_{ee}^{\rm exper}(i) \, \sigma_{\rm \mu\mu}^{\rm exper}(j-15)}{L(i) \, L(j-15)} \, V_{\rm L}(i,j-15) + \delta(i,j-15) \, \frac{\partial \sigma_{ee}^{\rm theor}}{\partial W}(i) \, \frac{\partial \sigma_{\rm \mu\mu}^{\rm theor}}{\partial W}(i) \, (\Delta W(i))^2 & \displaystyle \textcircled{\footnotesize{2}} \\
\displaystyle \frac{\sigma_{ee}^{\rm exper}(j) \, \sigma_{\rm \mu\mu}^{\rm exper}(i-15)}{L(i-15) \, L(j)} V_{\rm L}(i-15,j) + \delta(i-15,j) \, \frac{\partial \sigma_{ee}^{\rm theor}}{\partial W}(j) \, \frac{\partial \sigma_{\rm \mu\mu}^{\rm theor}}{\partial W}(j) \, (\Delta W(j))^2 & \displaystyle \textcircled{\footnotesize{3}} \\
\displaystyle V_{\rm \mu\mu}(i-15,j-15) + \delta(i-15,j-15) \, \left(\frac{\partial \sigma_{\rm \mu\mu}^{\rm theor}}{\partial W}(i-15) \, \Delta W(i-15)\right)^2 & \displaystyle \textcircled{\footnotesize{4}} \\
\end{array}
\right.
\label{Equation: Vij}
\end{equation}
with
\begin{equation}
\left\{
\begin{array}{l}
\displaystyle \textcircled{\footnotesize{1}} \text{ denotes } i=1,2 \cdots 14,15, \  j=1,2 \cdots 14,15 \\
\displaystyle \textcircled{\footnotesize{2}} \text{ denotes } i=1,2 \cdots 14,15, \  j=16,17 \cdots 29,30 \\
\displaystyle \textcircled{\footnotesize{3}} \text{ denotes } i=16,17 \cdots 29,30, \  j=1,2 \cdots 14,15 \\
\displaystyle \textcircled{\footnotesize{4}} \text{ denotes } i=16,17 \cdots 29,30, \  j=16,17 \cdots 29,30. \\
\end{array}
\right.
\end{equation}
\begin{multicols}{2}

\noindent Here, $\sigma_{ee}^{\rm exper}$ and $\sigma_{ee}^{\rm theor}$ ($\sigma_{\rm \mu\mu}^{\rm exper}$ and $\sigma_{\rm \mu\mu}^{\rm theor}$) are the experimental measured and theoretical predicted cross sections of the process \eetoee ($\mu^+\mu^-$), $V_{ee}$ ($V_{\rm \mu\mu}$) is the covariance matrix of the measured cross sections of \eetoee ($\mu^+\mu^-$), $V_{\rm L}$ is the covariance matrix of the measured luminosities, $i$ and $j$ are the horizontal and vertical indices of the $30 \times 30$ covariance matrix $V$, $\delta$ is Kronecker delta function, and $\Delta W$ is the statistcal uncertainty of the CM energy as listed in Table \ref{Table: Basic information of data.}, whose systematic uncertainty, 0.033 MeV, will be taken into account by examining the changes of the fit result due to the changes of the CM energies by 0.033 MeV.

\subsection{Simultaneous fit and parameter transformation}
\label{Subsection: Simultaneous Fit and Parameter Transformation}
By minimizing the global $\chi^2$ function, the simultaneous fit to the measured, CM energy dependent cross sections of the processes \eetoee and \eetomumu is carried out. In the fit, the following six parameters $M$, $\EwMEwDTw$, $\EwMMwDTw$, $S_{\rm W}$, $F_{ee}$ and $F_{\mu\mu}$ are float, while $\sigma^{\rm C}(W)$ and $R^{\rm FSR}(W)$ are expressed as piecewise linear interpolation functions based on hundreds of pairs of ($W$,$\sigma^{\rm C}$) and ($W$,$R^{\rm FSR}$) values obtained with the {\sc Babayaga-3.5} generator. The resultant fit curves are shown in Fig.~\ref{Figure: Simultaneous fit result of eetoee and eetomumu.}, and the corresponding fit quality is \ $\chi^2_{\rm min}/ndf\approx 23.0/24 \approx 1.0$, where $\chi^2_{\rm min}$ and $ndf$ are the minimized global chisquare and the number of degrees of freedom, respectively.

\end{multicols}
\begin{figurehere}
\centering
\includegraphics[width=0.85 \textwidth]{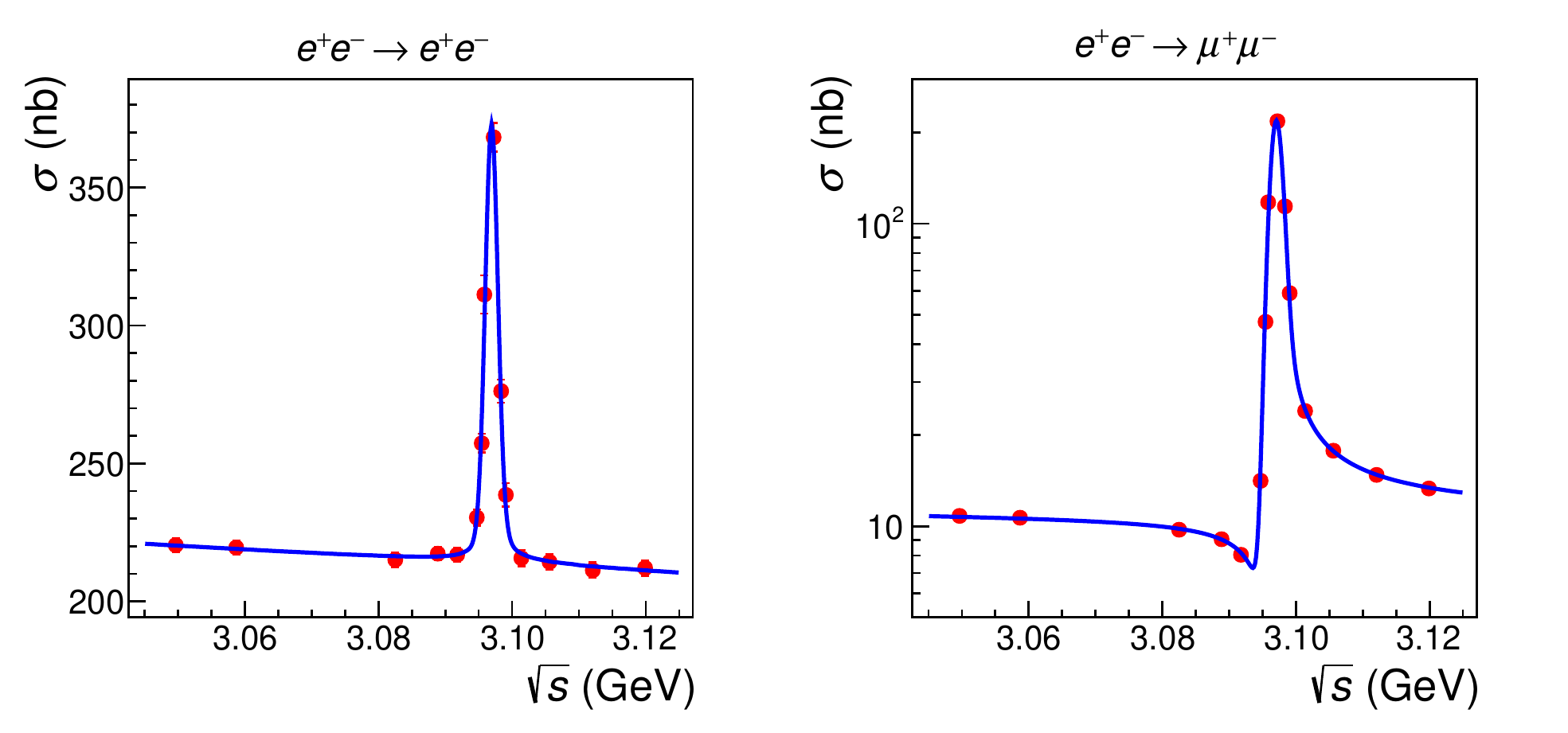}
\caption{\small Simultaneous fit to the measured, CM energy dependent
  cross sections of the processes \eetoee and \eetomumuwos. For clear
  display of the interference effect [denoted by $\sigma_{\rm 0}^{\rm
    I}$ in Eq. (\ref{Equation: ISR}) and mainly illustrated by the
  small dip in front of the peak], the plot for the process \eetomumu
  is drawn with a logarithmic vertical axis, while the plot for \eetoee
  is drawn with a linear vertical axis, because in this process the
  interference effect is less noticeable due to the existence of the
  scattering channel. In the plots, the red points with error bars and
  blue curves are from data and fitting, respectively.}
\label{Figure: Simultaneous fit result of eetoee and eetomumu.}
\end{figurehere}

\begin{table}[!h]
\captionsetup{width=0.725\textwidth}
\caption{\small Comparison of the \jpsi decay widths obtained in this work with those from other experiments, the PDG and the BESIII work using the ISR return technique.}
\centering
\begin{tabular*}{0.725\textwidth}{@{\extracolsep{\fill}}lccccc}
\hline
Collab. & Method & Year & $\Gamma_{\rm tot}$ (keV) & $\Gamma_{ll}$ (keV) & Ref.\\
\hline
BaBar & ISR return & 2004 & 94.7$\pm$4.4 & 5.61$\pm$0.21 & \cite{BaBar} \\
CLEO & ISR return & 2006 & 96.1$\pm$3.2 & 5.71$\pm$0.16 & \cite{CLEO} \\
KEDR & ES & 2010 & 94.1$\pm$2.7 & 5.59$\pm$0.12 & \cite{KEDR} \\
BESIII & ISR return & 2016 & --- & 5.58$\pm$0.09 & \cite{BESIII} \\
KEDR & ES & 2018 & 92.5$\pm$2.0 & 5.55$\pm$0.11 & \cite{KEDR2} \\
PDG & --- & 2020 & 92.9$\pm$2.8 & 5.53$\pm$0.10 & \cite{PDG} \\
This work & ES & 2021 & 93.0$\pm$2.1 & 5.56$\pm$0.11 & --- \\
\hline
\end{tabular*}
\label{Table: Comparison of this work with result from other experiments and with the PDG values.}
\end{table}

\begin{multicols}{2}

The fit result of $\EwMEwDTw$ and $\EwMMwDTw$ are ($0.346 \pm 0.009$) and ($0.335 \pm 0.006$) keV with the covariance and correlation coefficient between them as 0.000046 keV$^2$ and 0.83, respectively. Taking into account the correlation term, we evaluate $\EwDMw$ to be $1.031\pm0.015$, which is consistent with the expectation of lepton universality within about 2$\sigma$. Besides, $S_{\rm W}$ is fitted to be ($0.916\pm0.018$) MeV, which is consistent with the designed energy spread of the BEPCII collider, and $F_{ee}$ and $F_{\mu\mu}$ are fitted to be $0.995\pm0.009$ and $1.015\pm0.011$, respectively, which are compatible with the precision levels of the {\sc Babayaga-3.5} generator within uncertainties. 

Assuming lepton universality, $\Ew = \Mw$, and referring to $\Ew$ and $\Mw$ as $\Lw$, $\EwMEwDTw$ and $\EwMMwDTw$ are combined and referred to as $\LwMLwDTw$. The resultant $\LwMLwDTw$ is ($0.332 \pm 0.006$) keV, which is smaller than the individual values of $\EwMEwDTw$
and $\EwMMwDTw$ because of the correlation between the latter two.

Combining the resultant $\LwMLwDTw$ with the branching ratio of the \jpsi leptonic decay measured by BESIII in 2013, $\Bjpsitoll = \LwDTw = (5.978 \pm 0.040)$
\%~\cite{BESIII --- Branch ratio of Jpsi to ll}, $\Tw$
and $\Lw$ are determined to be ($93.0 \pm 2.1$) and ($5.56 \pm 0.11$) keV, respectively.

As mentioned previously, the impact of the systematic uncertainty (0.033 MeV) of the CM energies requires additional consideration. 
By increasing and decreasing the CM energies by 0.033 MeV, we repeat the entire simultaneous fit process twice, the relative changes of the results are less than 0.1\%, and are neglected.

\subsection{Result and comparison}
The results obtained in this work are summarized as follows:

\begin{align*}
& \EwMEwDTw = (0.346 \pm 0.009) \text{ keV,} \\
& \EwMMwDTw = (0.335 \pm 0.006) \text{ keV,} \\
& \LwMLwDTw = (0.332 \pm 0.006) \text{ keV,} \\
& \EwDMw = 1.031 \pm 0.015 \text{,} \\
& \Tw = (93.0 \pm 2.1) \text{ keV,} \\
& \Lw = (5.56 \pm 0.11) \text{ keV.}
\end{align*}

\noindent The uncertainties quoted here are total uncertainties, which are obtained with all the statistical and systematic uncertainties of the input quantities taken into consideration.

The result of $\EwDMw$ is consistent with and more precise than the result ($1.002 \pm 0.025$) given by KEDR with the same method~\cite{KEDR}. It is also in agreement with but less precise than the previous BESIII result ($1.0017 \pm 0.0037$) obtained with a different approach~\cite{BESIII --- Branch ratio of Jpsi to ll}. Table~\ref{Table: Comparison of this work with result from other experiments and with the PDG values.} shows a comparison of the $\Tw$ and $\Lw$ obtained in this work with those from other works and the PDG. The results given by this work agree with all other results; they come up to a new precision level, together with previous results obtained with the ISR return technique at BESIII and ES method at KEDR.

\section{Summary}
Based on the data samples collected with the BESIII detector at fifteen CM energy points in the vicinity of the \jpsi resonance, the cross sections of the processes \eetoee and \eetomumu are measured and summarized in Tables \ref{Table: Measured cross sections and related input quantities of the process eetoee at all individual CM energy points.} and \ref{Table: Measured cross sections and related input quantities of the process eetomumu at all individual CM energy points.}, respectively. By performing a simultaneous fit of the cross sections of the two processes as functions of the center-of-mass energy, $\EwMEwDTw$ and $\EwMMwDTw$ of the \jpsi resonance are determined to be ($0.346 \pm 0.009$) and ($0.335 \pm 0.006$) keV, respectively.

Using the obtained $\EwMEwDTw$ and $\EwMMwDTw$, the ratio of the \jpsi leptonic decay widths, $\EwDMw$, is evaluated to be $1.031 \pm 0.015$, which agrees with the expectation of lepton universality within about two standard deviations. Under the assumption of lepton universality and combining with $\Bjpsitoll$ measured by  BESIII~\cite{BESIII --- Branch ratio of Jpsi to ll}, the total and leptonic decay widths of the \jpsi resonance, $\Tw$ and $\Lw$, are determined to be ($93.0 \pm 2.1$) and ($5.56 \pm 0.11$) keV, respectively. These results are consistent with the previous results and reach the world leading level.

\section{Acknowledgments}
The BESIII collaboration thanks the staff of BEPCII and the IHEP computing center for their strong support. This work is supported in part by National Key R\&D Program of China under Contracts Nos.  2020YFA0406400, 2020YFA0406300; National Natural Science Foundation of China (NSFC) under Contracts Nos. 11275211, 11475090, 11335008, 11635010, 11735014, 11835012, 11935015, 11935016, 11935018, 11961141012, 12022510, 12025502, 12035009, 12035013, 12122509, 12192260, 12192261, 12192262, 12192263, 12192264, 12192265; the Chinese Academy of Sciences (CAS) Large-Scale Scientific Facility Program; Joint Large-Scale Scientific Facility Funds of the NSFC and CAS under Contract No. U1832207; 100 Talents Program of CAS; The Institute of Nuclear and Particle Physics (INPAC) and Shanghai Key Laboratory for Particle Physics and Cosmology; ERC under Contract No. 758462; European Union's Horizon 2020 research and innovation programme under Marie Sklodowska-Curie grant agreement under Contract No. 894790; German Research Foundation DFG under Contracts Nos. 443159800, Collaborative Research Center CRC 1044, FOR 2359, GRK 2149; Istituto Nazionale di Fisica Nucleare, Italy; Ministry of Development of Turkey under Contract No. DPT2006K-120470; National Science and Technology fund; National Science Research and Innovation Fund (NSRF) via the Program Management Unit for Human Resources \& Institutional Development, Research and Innovation under Contract No. B16F640076; STFC (United Kingdom); Suranaree University of Technology (SUT), Thailand Science Research and Innovation (TSRI), and National Science Research and Innovation Fund (NSRF) under Contract No. 160355; The Royal Society, UK under Contracts Nos. DH140054, DH160214; The Swedish Research Council; U. S. Department of Energy under Contract No. DE-FG02-05ER41374

\end{multicols}

\end{document}